\documentstyle[12pt]{article}
\begin{document}

\title{The newest release of the Ortocartan set of programs
for algebraic calculations in relativity}

\author{Andrzej Krasi\'nski\thanks{The implementation of the programs in Codemist
Standard Lisp described in this paper was supported by the Polish Research
Committee grant no 2 P03B 060 17.}}

\date {}

\maketitle

\centerline{N. Copernicus Astronomical Center, Polish Academy of Sciences }

\centerline{Bartycka 18, 00 716 Warszawa, Poland}

\centerline{email: akr@camk.edu.pl}

\begin{abstract}
The program Ortocartan for algebraic calculations in relativity has just been
implemented in the Codemist Standard Lisp and can now be used under the Windows
98 and Linux operating systems. The paper describes the new facilities and
subprograms that have been implemented since the previous release in 1992.
These are: the possibility to write the output as Latex input code and as
Ortocartan's input code, the calculation of the Ellis evolution equations for
the kinematic tensors of flow, the calculation of the curvature tensors from
given (torsion-free) connection coefficients in a manifold of arbitrary
dimension, the calculation of the lagrangian from a given metric by the
Landau-Lifshitz method, the calculation of the Euler-Lagrange equations from a
given lagrangian (only for sets of ordinary differential equations) and the
calculation of first integrals of sets of ordinary differential equations of
second order (the first integrals are assumed to be polynomials of second
degree in the first derivatives of the functions).

\end{abstract}

\section{The motivation}

Several programs for algebraic calculations are available on the market today,
and some of them include packages for relativity. Still, it seems that not all
possibilities in applying such programs are known to the users and to the
authors of those systems. The newest developments in Ortocartan include a few
facilities that are not yet in general use. It is hoped that the present paper
will demonstrate what is still possible beyond the standard applications.

\section{Calculating the curvature tensors from a given metric}

This has been the main application of the program Ortocartan ever since its
first release in 1978. It was described in several publications, most recently
in Ref. 1, so it will be recalled only very briefly. All details can be found
in Ref. 1, in the papers cited there, and in the latest release of the user's
manual [2].

The main program Ortocartan takes an orthonormal tetrad of Cartan forms as its
input data, and calculates all quantities that appear in the usual calculations
in relativity: the determinant of this tetrad, the components of the inverse
tetrad, the metric tensor and its inverse, the Ricci rotation coefficients, the
Christoffel symbols, the Riemann, Ricci, Einstein and Weyl tensors, the scalar
curvature. For the tensors calculated along the way it finds the (orthonormal)
tetrad components and the coordinate components with each index raised or
lowered as the user wishes. The user can make all kinds of substitutions,
including those by pattern-matching (for an example of the latter see sec. 8).
As an example, the input data for a very simple application of Ortocartan
(calculating the curvature tensors for the spherically symmetric metric in the
curvature coordinates) is shown below.

This example, and all the other examples shown in this paper, were chosen
trivial for illustrative purposes, so that the readers can easily see what
happens and how. In actual research, all the programs were successfully applied
also to very complicated examples.

\bigskip

\begin{verbatim}
(ortocartan '(
 (SPHERICAL METRIC IN STANDARD FORM)
 (functions  mu(t r)  nu(t r))
 (coordinates t r theta phi)
 (ematrix  (exp nu)  0  0  0  0  (exp mu) 0  0  0  0  r
             0  0  0  0  (r *(sin theta)))
 (stop after ricci)
 (rmargin 60)
                   ))
\end{verbatim}

\bigskip

\noindent (The command (rmargin 60) will fit the linelength of the output shown
in the next section to the present text.)

As before, the Ortocartan package includes the program Calculate that can carry
out simple algebraic operations defined by the user. This has been described in
the earlier publications, too, so again it is just recalled by the example
below, in which it calculates the derivative by $x$ of a complicated expression
that includes constants and functions. The input data is:

\bigskip

\begin{verbatim}
(calculate '(
    (print example)
    (constants a b c d)
    (coordinates x)
    (functions f (x))
    (operation (deriv x (a ^ (b ^ ((der x f) ^ (c ^ d))))))
        ))
\end{verbatim}

\bigskip

\noindent and the output is\footnote{Each piece of output shown further is a
part of a disk-output from the appropriate program that was inserted verbatim
in the latex code of this paper.}:

\bigskip

{\samepage
\begin{verbatim}
(print example)

(I UNDERSTAND YOU REQUEST THE FOLLOWING EXPRESSION TO BE SIMPLIFIED)

                        d
                       c
                   f,
                     x
                  b
>       deriv (x,a       )
\end{verbatim} }
\begin{verbatim}

THE RESULT IS

                     d
                    c        d
                f,          c                                       d
                  x
               b        f,      d                            - 1 + c
                          x
>   result  = a        b       c  log (a) log (b) f,     f,
          1                                         x x    x

(I REALLY LIKED THIS ! CAN I HAVE MORE ? PLEASE ?!?)

 END OF WORK
 (RUN TIME = 50 msec)
\end{verbatim}

\bigskip

\noindent This is a fictitious example, not taken from any actual application,
and was intended to demonstrate the power of our differentiating and printing
procedures.

\section{Output in the form of Latex input and in the form of Ortocartan's input}

The simple call to Ortocartan shown above would result in printing all the
formulae in the standard mathematical format. One item of the output is shown
below (the tetrad $R_{00}$ component of the Ricci tensor):

\bigskip

{\samepage
\begin{verbatim}
                  -1
>   ricci    = 2 r   exp (- 2 mu) nu,   + exp (- 2 mu) nu,
         0 0                         r
\end{verbatim}
}
 {\samepage
\begin{verbatim}
          2
>           + exp (- 2 mu) nu,     - exp (- 2 mu) mu,   nu,
        r                     r r                    r
\end{verbatim}
}
 {\samepage
\begin{verbatim}
                               2
>          - exp (- 2 nu) mu,    - exp (- 2 nu) mu,     +
        r                    t                     t t
\end{verbatim}
}
 {\samepage
\begin{verbatim}

>       exp (- 2 nu) mu,   nu,
                        t     t
\end{verbatim}
}

\bigskip

\noindent This can be either just viewed on the screen or stored on a disk and
then printed. However, if a formula like the one above is to be inserted in a
text for publication, then one may avoid retyping it. The same formula can be
obtained in the form of Latex input code if one inserts the following item
anywhere between the arguments of the function Ortocartan (but after the title
(SPHERICAL METRIC...)):

\bigskip

\begin{verbatim}
(output for latex)
\end{verbatim}

\bigskip

\noindent The same component of the Ricci tensor will then appear on output in
the following form

\bigskip

\begin{verbatim}
\begin{equation}
ricci_{00} = 2r^{-1}\exp (- 2{\mu}){\nu},_{r} + \exp (- 2{\mu}){{\nu},_{r}}^{2}
$$

$$ + \exp (- 2{\mu}){\nu},_{r r} - \exp (- 2{\mu}){\mu},_{r}{\nu},_{r} - \exp
(- 2{\nu}){{\mu},_{t}}^{2} - \exp (- 2{\nu}) $$

$$ {\mu},_{t t} + \exp (- 2{\nu}){\mu},_{t}{\nu},_{t}
\end{equation}
\end{verbatim}

\bigskip

\noindent Note that Ortocartan has recognized the Greek letters and replaced
them with the appropriate Latex constructs. Then the user can add his/her
favourite Latex preamble to this and run Latex on the result obtaining the
following output:

\bigskip

\begin{equation}
ricci_{00} = 2r^{-1}\exp (- 2{\mu}){\nu},_{r} + \exp (- 2{\mu}){{\nu},_{r}}^{2}
$$

$$ + \exp (- 2{\mu}){\nu},_{r r} - \exp (- 2{\mu}){\mu},_{r}{\nu},_{r} - \exp
(- 2{\nu}){{\mu},_{t}}^{2} - \exp (- 2{\nu}) $$

$$ {\mu},_{t t} + \exp (- 2{\nu}){\mu},_{t}{\nu},_{t}
\end{equation}

\bigskip

The \verb+exp(f)+ can be very easily replaced by ${\rm e}^f$ using the
substitutions. The exponential function is represented as above in order to
give the user a greater freedom: the symbol "e" can be used for anything the
user wishes and is not reserved to mean only the base of natural logarithms.

If the output or some part of it is to be used as input for one of the programs
of the Ortocartan set, then the user can avoid rewriting again. It is then
enough to insert the following instead of the "(output for latex)" line:

\bigskip

\begin{verbatim}
(output for input)
\end{verbatim}

\bigskip

\noindent The whole output will then be written in the Ortocartan input
notation and any part of it can be inserted in the actual input just by cutting
and pasting. The same component of the Ricci tensor that was shown above would
then appear in the following form:

\bigskip

\begin{verbatim}
ricci (0 0) = ((2 * (r ^ -1) * (exp (-2 * mu)) * (der r nu)) + ((exp (-2 * mu
)) * ((der r nu) ^ 2)) + ((exp (-2 * mu)) * (der r r nu)) + (-1 * (exp (-2 *
mu)) * (der r mu) * (der r nu)) + (-1 * (exp (-2 * nu)) * ((der t mu) ^ 2)) +
(-1 * (exp (-2 * nu)) * (der t t mu)) + ((exp (-2 * nu)) * (der t mu) * (der t
nu)))
\end{verbatim}

\bigskip

\noindent The automatically generated output in the input format has the
tendency to use more parentheses than are necessary. The algebraically
equivalent input written by the user would not contain any of the unnecessary
parentheses because Ortocartan's input notation recognizes the usual hierarchy
of priorities among the algebraic operations -- see manual [2]. The additional
parentheses do not change this hierarchy.

The "(output for latex)" and "(output for input)" options are available for all
the other programs described below.

\section{The program Ellisevol.}

This program calculates all the quantities appearing in the evolution equations
of the kinematical tensors of fluid flow, as defined by Ellis [3]. Since all
these equations are consequences of the Ricci identity $u_{\alpha;\beta \gamma}
- u_{\alpha; \gamma \beta} = u_{\mu}{R^{\mu}}_{\alpha \beta \gamma}$, they will
be fulfilled identically in most cases. However, they may fail to be
identically fulfilled when one makes assumptions about separate parts of the
flow, e.g. if one assumes that the shear is zero. As is well known, such
assumptions have consequences for the other characteristics of the flow, and
the Ellis equations will show what the consequences are. Along the way, the
program calculates all those quantities that are calculated by the program
Ortocartan, and in addition the expansion, the acceleration, the shear tensor
and scalar, the rotation tensor and scalar, and the electric and magnetic parts
of the Weyl tensor with respect to the velocity vector.

Since the signature assumed in the calculation is $(+ - - -)$, the formulae may
differ from those given in textbooks, and so some of them are quoted below for
reference. The equations that the program verifies are the following.

The rotation constraint equations:

$$ \omega_{[\alpha \beta; \gamma]} + \dot{u}_{[\alpha; \gamma}u_{\beta]} +
\dot{u}_{[\alpha}\omega_{\beta \gamma]} = 0 $$

\noindent (square brackets on indices denote antisymmetrization, round brackets
on indices denote symmetrization).

The shear constraint equations:

$${h^{\alpha}}_{\beta}({\omega^{\beta \gamma}}_{;\gamma} - {\sigma^{\beta
\gamma}}_{;\gamma} + {2 \over 3} \theta^{;\beta}) - ({\omega^{\alpha}}_{\beta}
+ {\sigma^{\alpha}}_{\beta})\dot{u}^{\beta} = 0, $$

\noindent where ${h^{\alpha}}_{\beta} = {\delta^{\alpha}}_{\beta} -
u^{\alpha}u_{\beta}$ is the projection tensor.

The rotation evolution equations:

$${h_{\alpha}}^{\gamma} {h_{\beta}}^{\delta}\dot{\omega}_{\gamma \delta} -
{h_{\alpha}}^{\gamma} {h_{\beta}}^{\delta} \dot{u}_{[\gamma; \delta]} + 2
\sigma_{\delta [\alpha} {\omega^{\delta}}_{\beta]} + {2 \over 3} \theta
\omega_{\alpha \beta} = 0. $$

The Raychaudhuri equation:

$$\dot{\theta} + {1 \over 3}\theta^2 - {\dot{u}^{\alpha}}_{; \alpha} +
\sigma^{\alpha \beta} \sigma_{\alpha \beta} - \omega^{\alpha \beta}
\omega_{\alpha \beta} + R_{\alpha \beta}u^{\alpha} u^{\beta} = 0. $$

The (coordinate) electric components of the Weyl tensor:

$$E_{\alpha \beta} = C_{\alpha \rho \beta \sigma} u^{\rho} u^{\sigma} \equiv
E_{\beta \alpha}. $$

The shear evolution equations:

$${h_{\alpha}}^{\gamma} {h_{\beta}}^{\delta}\dot{\sigma}_{\gamma \delta} -
{h_{\alpha}}^{\gamma} {h_{\beta}}^{\delta} \dot{u}_{(\gamma; \delta)} +
\dot{u}_{\alpha} \dot{u}_{\beta} + \omega_{\alpha \gamma}
{\omega^{\gamma}}_{\beta} + \sigma_{\alpha \gamma} {\sigma^{\gamma}}_{\beta} +
{2 \over 3} \theta \sigma_{\alpha \beta} $$

$$+ {1 \over 3} h_{\alpha \beta}[2(\omega^2 - \sigma^2) +
{{\dot{u}^{\gamma}}}_{; \gamma}] + E_{\alpha \beta} = 0. $$

The magnetic components of the Weyl tensor:

$$ H_{\alpha \beta} = {1 \over 2} {\sqrt {- g}}\varepsilon_{\alpha \gamma \mu
\nu} {C^{\mu \nu}}_{\beta \delta} u^{\gamma} u^{\delta} \equiv H_{\beta
\alpha}, $$

\noindent where $\varepsilon_{\alpha \gamma \mu \nu}$ is the Levi-Civita
symbol.

The "magnetic constraint" equations:

$$ 2\dot{u}_{(\alpha}w_{\beta)} - {\sqrt {- g}}{h_{\alpha}}^{\gamma}
{h_{\beta}}^{\delta} ({\omega_{(\gamma}}^{\mu ; \nu} + {\sigma_{(\gamma}}^{\mu
; \nu})\varepsilon_{\delta) \rho \mu \nu} u^{\rho} = H_{\alpha \beta}, $$

\noindent where $w_{\beta}$ is the rotation vector field defined by:

$$w^{\alpha} = {1 \over {2 \sqrt{- g}}} \varepsilon^{\alpha \beta \gamma
\delta} u_{\beta} \omega_{\gamma \delta}. $$

As an example, let us consider the application of this program to the metric of
Lanczos [4].

$$ ds^2 = (dt + Crd\varphi)^2 - \psi d\varphi^2 - {1 \over 4} {\rm e}^{- r}
dr^2/\psi - {\rm e}^{- r}dz^2, $$

\noindent where $C =$ const and

$$ \psi = (C^2 r + \Lambda - \Lambda {\rm e}^{- r}).$$

\bigskip

This is a stationary cylindrically symmetric solution of Einstein's equations
with a rotating dust source and with a nonvanishing cosmological constant
$\Lambda$. The coordinates used in the metric shown above are comoving and the
velocity vector field of the dust is one of the orthonormal tetrad vectors,
hence the tetrad components of the velocity field are \verb+(1 0 0 0)+. Since
this is a solution of Einstein's equations, this vector field is uniquely
determined by the metric, and so, as expected, all the constraint and evolution
equations will be identities. However, the acceleration (= 0), rotation,
expansion (= 0), and shear (= 0) are all calculated, along with the electric
and magnetic parts of the Weyl tensor.

The input data is here:

\bigskip

\begin{verbatim}
 (setq !*lower nil)
 (ellisevol'(
    (LANCZOS METRIC)
    (coordinates t phi r z)
    (velocity 1 0 0 0)
    (constants C Lambda)
    (symbols psi = (C ^ 2 * r + Lambda - Lambda * (exp (-  r))) )
    (ematrix  1  (C * r)  0  0
             0  ((C ^ 2 * r + Lambda - Lambda * (exp (-  r)))^ (1 2 ))
         0 0 0 0 ((1 2) * (exp ((-1 2) * r)) * (C ^ 2 * r + Lambda
                   - Lambda * (exp (- r))) ^ (-1 2))  0
             0  0  0  (exp (- (1 2) * r)))
 (substitutions (C ^ 2 * r + Lambda - Lambda * (exp (-  r))) = psi )
   (dont print messages)
   (tensors einstein)
          ))
(setq !*lower t)
\end{verbatim}

\bigskip

\noindent The first line of the input tells the system to treat the lower-case
letters and their corresponding capitals as different symbols, the last line
reverses this command. The command "(dont print messages)" tells the system not
to print the messages about unsuccessful attempts to carry out the substitution
specified in the previous line. Without this command, the program would write a
message every time when there was no opportunity for this substitution in a
given formula.

The most important parts of the output are given below. Each tensor is printed
with its unique name to facilitate substitutions. Thus the covariant rotation
tensor $\omega_{\alpha \beta}$ has the name "rotdd", while the mixed rotation
tensor ${\omega_{\alpha}}^{\beta}$ has the name "rotdu", and similarly for
shear.

\bigskip

\begin{verbatim}
ACCELERATION = 0

\end{verbatim}
{\samepage
\begin{verbatim}
>   rotdd    = (1/2) C
         1 2
\end{verbatim}
  }
\begin{verbatim}
(rotdd calculated)
(TIME = 1060 msec)

(rotdd completed)
(TIME = 1070 msec)
\end{verbatim}
{\samepage
\begin{verbatim}
           2
>   rotdu    = - 2 C exp (r) psi
         1
\end{verbatim}
 }
 {\samepage
\begin{verbatim}
           0            2      -1
>   rotdu    = - (1/2) C  r psi
         2
\end{verbatim}
 }
 {\samepage
\begin{verbatim}
           1              -1
>   rotdu    = (1/2) C psi
         2
\end{verbatim}
 }
\begin{verbatim}
(rotdu completed)
(TIME = 1100 msec)
\end{verbatim}
 {\samepage
\begin{verbatim}
                        2
>   ROTATION SQUARED = C  exp (r)
\end{verbatim}
 }
\begin{verbatim}
(ROTATION SCALAR calculated)
(TIME = 1100 msec)

>   EXPANSION SCALAR = 0

(EXPANSION SCALAR calculated)
(TIME = 1150 msec)

(sheardd calculated)
(TIME = 1160 msec)

SHEAR = 0

(ALL THE ROTATION CONSTRAINTS ARE FULFILLED IDENTICALLY)
(ROTATION CONSTRAINTS calculated)
(TIME = 1160 msec)

(ALL THE SHEAR CONSTRAINTS ARE FULFILLED IDENTICALLY)
(SHEAR CONSTRAINTS calculated)
(TIME = 2500 msec)

(ALL THE ROTATION EVOLUTION EQUATIONS ARE FULFILLED IDENTICALLY)
(ROTATION EVOLUTION EQUATIONS calculated)
(TIME = 2800 msec)

>   RAYCHAUDHURI EQUATION = 0

(RAYCHAUDHURI EQUATION calculated)
(TIME = 3150 msec)
\end{verbatim}
 {\samepage
\begin{verbatim}
                         2
>   elweyl    = - (1/3) C  exp (r) psi
          1 1
\end{verbatim}
 }
 {\samepage
\begin{verbatim}
                          2    -1
>   elweyl    = - (1/12) C  psi
          2 2
\end{verbatim}
 }
 {\samepage
\begin{verbatim}
                       2
>   elweyl    = (2/3) C
          3 3
\end{verbatim}
 }
\begin{verbatim}
(elweyl calculated)
(TIME = 3310 msec)

(ALL THE SHEAR EVOLUTION EQUATIONS ARE FULFILLED IDENTICALLY)
(SHEAR EVOLUTION EQUATIONS calculated)
(TIME = 3820 msec)
\end{verbatim}
 {\samepage
\begin{verbatim}
>   magweyl    = - (1/2) C
           2 3
\end{verbatim}
 }
\begin{verbatim}
(magweyl calculated)
(TIME = 3890 msec)

(ALL THE MAGNETIC CONSTRAINTS ARE FULFILLED IDENTICALLY)
(magcons calculated)
(TIME = 5170 msec)

(I REALLY LIKED THIS! CAN I HAVE MORE ?   PLEASE ?!?)

END OF WORK
(RUN TIME = 5170 msec)
\end{verbatim}

\bigskip
\section{The program Curvature.}

This program calculates the curvature tensor from given connection coefficients
in any number of dimensions. The connection coefficients are assumed symmetric
(i.e. torsion-free), but need not be metrical. The program was written for one
special application, and hence the assumption of zero torsion; it can be
removed without any difficulty.

\section{The program Landlagr.}

This program calculates the reduced lagrangian for the Einstein equations by
the Landau--Lifshitz method [5]. This is the Hilbert lagrangian with the
divergences containing second derivatives of the metric already removed. In
short, this (noncovariant) lagrangian is obtained by deleting from the scalar
curvature those terms in which the Christoffel symbols are differentiated, and
taking the remaining part with the reverse sign.

As is well-known, the Euler-Lagrange equations derived from a variational
principle in which the generality of the metric was limited by various
assumptions (e.g. about symmetry) are not always equivalent to the Einstein
equations. It may happen that they will have nothing to do with the Einstein
equations; this is known, for example, to occur for certain Bianchi-type models
[6]. Therefore, the user must make sure, when using the program "landlagr",
that the situation is appropriate for applying the lagrangian methods.

\section{The program Eulagr.}

This program calculates the Euler-Lagrange equations starting from a lagrangian
specified by the user. It is assumed that the resulting E-L equations will be
ordinary differential equations (i.e. that there is only one independent
variable in the lagrangian action integral).

As an example, the program will derive the Newtonian equations of motion for a
point particle of mass $m$ in the cartesian coordinates $\{x, y, z\}$ from the
lagrangian

$$ L = {1 \over 2} m ({\dot {x}}^2 + {\dot {y}}^2 + {\dot {z}}^2) - V(x, y, z),
$$

\noindent where $V$ is a potential and $x(t), y(t), z(t)$ are the equations of
a trajectory of the particle. The input data is:

\bigskip

\begin{verbatim}
(setq !*lower nil)
(eulagr '(
   (The lagrangian for the Newtonian equations of motion
    in 3 dimensions)
   (constants m)
   (parameter t)
   (functions x(t) y(t) z(t) V(x y z) )
   (variables x y z)
   (lagrangian ((1 2) * m * ((der t x) ^ 2 + (der t y) ^ 2
      + (der t z) ^ 2) - V))
        ))
(setq !*lower t)
\end{verbatim}

\bigskip

\noindent and the results are:

\bigskip

\begin{verbatim}
(The lagrangian for the Newtonian equations of motion in 3 dimensions)
\end{verbatim}
 {\samepage
\begin{verbatim}
                                   2               2               2
>   lagrangian = - V + (1/2) m x,    + (1/2) m y,    + (1/2) m z,
                                 t               t               t
\end{verbatim}
}
{\samepage
\begin{verbatim}
(THIS IS THE VARIATIONAL DERIVATIVE BY x)

>   eulagr  = m x,     + V,
          0       t t      x
\end{verbatim}
}
{\samepage
\begin{verbatim}
(THIS IS THE VARIATIONAL DERIVATIVE BY y)

>   eulagr  = m y,     + V,
          1       t t      y
\end{verbatim}
}
{\samepage
\begin{verbatim}
(THIS IS THE VARIATIONAL DERIVATIVE BY z)

>   eulagr  = m z,     + V,
          2       t t      z
\end{verbatim}
 }
\begin{verbatim}
(I REALLY LIKED THIS! CAN I HAVE MORE ?   PLEASE ?!?)

END OF WORK
(RUN TIME = 100 msec)
\end{verbatim}

\bigskip

All the other programs described in this paper are, from the programmer's point
of view, rather similar to each other. The various programs are simply
different sets of the same basic algebraic operations. The program Eulagr is
somewhat exceptional, since, when calculating variational derivatives, the
functional expressions $\dot{x}^i, i = 1, 2, ..., n$ have to be treated as
independent variables at a certain stage. The program generates names for these
variables, adds them to the list of independent variables for differentiation,
replaces the $\dot{x}^i$ by the new variables, then calculates the derivatives
$\partial L / \partial \dot{x}^i$ and $\partial L / \partial x^i$, then
replaces the new variables back by $\dot{x}^i$ in the resulting expressions,
and finally uses these results to calculate ${{\rm d} \over {{\rm d} t}}
(\partial L /
\partial \dot{x}^i)$. All this happens automatically, and the user need not
worry about any details of it. Ortocartan's differentiating and substituting
subprograms are flexible enough to handle this.

\section{The program Squint.}

This program verifies whether a given expression is a first integral of a given
set of ordinary differential equations. The program was written for a specific
application, therefore it is rather limited in its abilities. It is assumed
that the (hypothetical or actual) first integral is a polynomial of first or
second degree in the first derivatives of the functions that should obey the
set of equations. It is also assumed that the equations in the set are all of
second or first order and that they have the form:

$$ {f^i},_{tt} = \Psi^i(f^1, ..., f^n, {f^1},_t, ..., {f^n},_t) $$

\noindent i.e. that they are algebraically solved with respect to the highest
derivatives. "Squint" is an abbreviation for "square integral".

In order to make the result easy to verify, the program "squint" will be used
to find a first integral of the equations found in the previous example. We
shall at first pretend that we do not know what the integral $I$ should be and
will assume that it is a general polynomial of second degree in the first
derivatives by $t$ of the functions $x(t)$, $y(t)$ and $z(t)$, thus $I =
Q_{ij}{{{\rm d}x^i} \over {{\rm d} t}}{{{\rm d}x^j} \over {{\rm d} t}} + L_i
{{{\rm d}x^i} \over {{\rm d} t}} + E$, where $Q_{ij}$, $L_i$ and $E$ are
unknown functions of the $x^i$. The argument "(markers M)" defines $M$ as a
symbolic variable that can be used to represent any expression. Thanks to this
facility the single equation (der t t M) = (- (der M V) / m) represents the 3
equations $d^2x^i/dt^2 = - \partial V /\partial x^i$ for $i = 1, 2, 3$
simultaneously. The "maineq" directs these substitutions to the main equation,
which is the explicit formula for the total derivative ${\rm d}I/{\rm d}t$.

The input data is:

\bigskip

\begin{verbatim}
(setq !*lower nil)
(squint'(
   (a first integral of the Newtonian equations of motion)
   (constants m)
   (parameter t)
   (functions x(t) y(t) z(t) V(x y z) Q11(x y z) Q12(x y z) Q13(x y z)
   Q22(x y z) Q23(x y z) Q33(x y z) L1(x y z) L2(x y z) L3(x y z)
   E(x y z) )
   (variables x y z)
(integral (Q11 * (der t x) ^ 2 + 2 * Q12 * (der t x) * (der t y)
    + 2 * Q13 * (der t x) * (der t z) + Q22 * (der t y) ^ 2
    + 2 * Q23 * (der t y) * (der t z) + Q33 * (der t z) ^ 2
    + L1 * (der t x) + L2 * (der t y) + L3 * (der t z) + E) )
(markers M)
   (substitutions
    maineq
(der t t M) = (- (der M V) / m)
     )
(dont print maineq)
          ))
(setq !*lower t)
\end{verbatim}

\bigskip

The output includes the integral $I$ printed in the mathematical format, the
"main equation" ${\rm d}I/{\rm d}t$ (the printing of ${\rm d}I/{\rm d}t$ was
suppressed by the command "(dont print maineq)"), the coefficients of all the
expressions ${{{\rm d}x^i} \over {{\rm d} t}} {{{\rm d}x^j} \over {{\rm d} t}}
{{{\rm d}x^k} \over {{\rm d} t}}$, ${{{\rm d}x^i} \over {{\rm d} t}} {{{\rm
d}x^j} \over {{\rm d} t}}$, ${{{\rm d}x^i} \over {{\rm d} t}}$, and the
remaining part of ${\rm d}I/{\rm d}t$ that does not contain any ${{{\rm d}x^i}
\over {{\rm d} t}}$. This will be done correctly only if the second derivatives
${{{\rm d}^2x^i} \over {{\rm d} t^2}}$ appearing in ${\rm d}I/{\rm d}t$ had
been replaced using the (substitutions ...), in this way use is made of the set
of equations for which $I$ is expected to be the first integral. If the set
consists of $n$ equations, then the number of coefficients printed will be ${1
\over 6}(n + 1)(n + 2)(n + 3)$, i.e. 20 for $n = 3$. When the explicit form of
$I$ is unknown, as in this example, the coefficients will be partial
differential equations that should determine the $Q_{ij}$, $L_i$ and $E$ (each
of the "equations" printed is a left-hand side of an equation that has zero on
the right-hand side).

Only a few pieces of the output are reproduced below.

\bigskip

\begin{verbatim}
(a first integral of the Newtonian equations of motion)
\end{verbatim}
{\samepage
\begin{verbatim}
                           2
>   integral = E + Q11 x,    + 2 Q12 x,   y,   + 2 Q13 x,   z,   + Q22
                         t             t    t            t    t
\end{verbatim}
}
{\samepage
\begin{verbatim}
            2                             2
>       y,    + 2 Q23 y,   z,   + Q33 z,    + L1 x,   + L2 y,   + L3 z,
          t             t    t          t          t         t
\end{verbatim}
}
{\samepage
\begin{verbatim}
>
        t
\end{verbatim}
}
{\samepage
\begin{verbatim}
                                     3
>   THIS IS THE COEFFICIENT OF   x,
                                   t
\end{verbatim}
}
{\samepage
\begin{verbatim}
>   equation  = Q11,
            1       x
\end{verbatim}
}
{\samepage
\begin{verbatim}
                                     2
>   THIS IS THE COEFFICIENT OF   x,    y,
                                   t     t
\end{verbatim}
}
{\samepage
\begin{verbatim}
>   equation  = 2 Q12,   + Q11,
            2         x        y
\end{verbatim}
}
\begin{verbatim}

..............................
 [part of the output deleted]
..............................

>   THESE ARE THE TERMS THAT ARE   FREE OF THE DERIVATIVES
\end{verbatim}
{\samepage
\begin{verbatim}
                    -1            -1            -1
>   equation   = - m   L1 V,   - m   L2 V,   - m   L3 V,
            20              x             y             z
\end{verbatim}
}
\begin{verbatim}
(I REALLY LIKED THIS! CAN I HAVE MORE ?   PLEASE ?!?)

END OF WORK
(RUN TIME = 1030 msec)
\end{verbatim}

\bigskip

Now let us substitute the well-known solution of these equations into the
data and see what happens.

\bigskip

\begin{verbatim}
(setq !*lower nil)
(squint'(
   (a first integral of the Newtonian equations of motion - the final result)
   (constants m)
   (parameter t)
   (functions x(t) y(t) z(t) V(x y z))
   (variables x y z)
(integral ((1 2) * m * ((der t x) ^ 2 + (der t y) ^ 2
                          + (der t z) ^ 2) + V) )
(markers M)
   (substitutions
    maineq
(der t t M) = (- (der M V) / m)
     )
     (dont print maineq)
          ))
(setq !*lower t)
\end{verbatim}

\bigskip

\noindent The result is:

\bigskip

\begin{verbatim}
(a first integral of the Newtonian equations of motion - the final result)
\end{verbatim}
 {\samepage
\begin{verbatim}

                               2               2               2
>   integral = V + (1/2) m x,    + (1/2) m y,    + (1/2) m z,
                             t               t               t
\end{verbatim}
 }
\begin{verbatim}

(THE FIRST INTEGRAL IS ALREADY MAXIMALLY SIMPLIFIED
AND IS EXPLICITLY CONSTANT)
>   maineq = 0

(I REALLY LIKED THIS! CAN I HAVE MORE ?   PLEASE ?!?)

END OF WORK
(RUN TIME = 130 msec)
\end{verbatim}

\bigskip

Similar programs for verifying first integrals can be written for any kind of
dependence of the first integral on the first derivatives of the functions.

\bigskip

\appendix

\section{How to obtain Ortocartan.}

The base language of this newest release of Ortocartan is Codemist Standard
Lisp. The latter is implemented in Windows 98 and in Linux, and so should be
usable for most computer users.

In order to use Ortocartan one must first buy the Codemist Standard Lisp. It
can be bought from:

\bigskip

     Professor John Fitch, Director

     CODEMIST Limited

     "Alta", Horsecombe Vale

     Combe Down

     BATH, Avon, BA2 5QR

     England

     phone and fax (44-1225) 837 430

     email jpffitch@maths.bath.ac.uk

\bigskip

Ortocartan is free of charge. If you wish to have it, just contact A.
Krasi\'nski, I will either email the programs to you or send you a diskette.
How to install Ortocartan when Lisp is already working is described in sec. 7
of the manual [2].

\bigskip

\centerline{\bf REFERENCES}

[1] A. Krasi\'nski, {\it Gen. Rel. Grav.} {\bf 25}, 165 (1993).

[2] A. Krasi\'nski, M. Perkowski, {\it The system Ortocartan - user's manual}.
Fifth edition, Warszawa 2000. A Latex file distributed by email or on
diskettes.

[3] G. F. R. Ellis, in {\it General relativity and cosmology. Proceedings of
the International School of Physics "Enrico Fermi", Course 47: General
Relativity and Cosmology.} Edited by R. K. Sachs, Academic Press, New York
1971.

[4] K. Lanczos, {\it Zeitschrift f\"{u}r Physik} {\bf 21}, 73
(1924); {\it Gen. Rel. Grav.} {\bf 29}, 363 (1997).

[5] L. D. Landau and E. M. Lifshitz, {\it Teoria polya}, 6th Russian edition.
Izdatelstvo "Nauka", Moskva 1973, sec, 93.

[6] M. A. H. MacCallum, in {\it General relativity. An Einstein centenary
survey.} Edited by S. W. Hawking and W. Israel. Cambridge University Press
1979, p. 552.

\end{document}